# Phase Transition Under Control: Toward Application-Oriented Luminescence Thermometry and Thermally Activated Emission

M. T. Abbas[1], M. Szymczak[1*], D. Szymanski[1], J. Zeler[2], M. Drozd[1], L. T. K Giang[3], L. Marciniak[1*]

[1] Institute of Low Temperature and Structure Research, Polish Academy of Sciences, Okólna 2, 50-422 Wrocław, Poland

[2] University of Wrocław, Faculty of Chemistry, 14. F Joliot Curie Street, 50-383 Wrocław, Poland

[3] Institute of Materials Science, Vietnam Academy of Science and Technology, Hanoi, Vietnam

*corresponding author: l.marciniak@intibs.pl, m.szymczak@intibs.pl

**Abstract**

Phase-transition-based luminescent thermometers are characterized by two inherent limitations: a narrow thermal operating range and the presence of a hysteresis loop in the thermometric parameter. In this work, we demonstrate that controlling the particle size of $LaGaO_3:Eu^{3+}$ phosphors enables significant enhancement of thermometric performance. Specifically, a reduction in grain size dispersion leads to an increase in relative thermal sensitivity and significantly narrows the hysteresis loop. As a result of this approach, the relative sensitivity was increased to 18.2% $K^{-1}$ for $LaGaO_3:Eu^{3+}$ synthesized via the solid-state method, compared to 3.0% $K^{-1}$ for the counterpart prepared using the Pechini method. Furthermore, we show that the intentional incorporation of $Al^{3+}$ and $Sc^{3+}$ co-dopant ions allows for continuous tuning of the structural phase transition temperature from 165 K for 15% $Al^{3+}$ to 491 K for 2% $Sc^{3+}$, without significantly affecting the low-temperature spectroscopic properties of $Eu^{3+}$ ions. This ability to shift the phase transition temperature in $LaGaO_3$ offers a practical route to modulate the thermal response range of the luminescent thermometer, enabling its adaptation to specific application requirements. The empirical relationship established in this study between the phase transition temperature and the ionic radius mismatch parameter provides a predictive tool for the rational design of phase-transition-based phosphors with tailored thermometric performance. The ability to systematically tune the phase transition temperature via ionic radius mismatch, together with enhanced thermometric performance resulting from reduced grain size dispersion, establishes a coherent strategy for the rational design of high-sensitivity, low-hysteresis thermal sensors. These results lay the groundwork for the development of application-specific luminescent thermometers with tailored operating ranges and improved reliability

**Introduction**

Undoubtedly, the most extensively studied class of luminescent thermometers comprises inorganic materials doped with lanthanide ions ($Ln^{3+}$), with particular emphasis on ratiometric Boltzmann-type thermometers[1–4]. These systems offer the advantage of a theoretically predictable calibration curve due to the well-defined energy levels of $Ln^{3+}$ ions. However, a key limitation lies in their relatively low sensitivity, which is inherently constrained by the energy gap between the thermally coupled excited states[2,5]. Recent studies have highlighted an important alternative - luminescent thermometers based on thermally induced first-order structural phase transitions[6–14]. In such systems, a phase transition significantly alters the local environment of the $Ln^{3+}$ ions, leading to pronounced changes in their spectroscopic properties. This mechanism can yield exceptionally high relative sensitivity values (up to 26.1 % $K^{-1}$)[13]. Nevertheless, two primary limitations of this approach remain: (1) the typically narrow thermal operating range, limited to temperatures near the phase transition, and (2) the presence of a hysteresis loop in the thermal dependence of the thermometric parameter – typically luminescence intensity ratio (LIR), which can affect temperature readout precision[7,13]. To address the first limitation, a co-doping strategy has recently been proposed, in which optically inactive ions with different ionic radii replace the cations occupied by $Ln^{3+}$ ions[8,14]. This induces structural strain within the lattice, effectively shifting the phase transition temperature and, consequently, modifying the thermal operating range of luminescent thermometer. However, due to the relatively small differences in ionic radii among rare-earth ions, the resulting shift is often limited. Additionally, such substitutions can be synthetically challenging or even unfeasible in certain host materials. In this work, we present an alternative approach involving cation substitution in the second coordination sphere, where significantly larger ionic radii differences can be exploited. The study focuses on a model system: $LaGaO_3$:$Eu^{3+}$ which undergoes of the structural first order phase transition from orthorhombic to trigonal phases above around 440 K[15–19]. In this compound, $Eu^{3+}$ ions substitute $La^{3+}$ sites, while the structural modification is introduced through partial replacement of $Ga^{3+}$ ($R$=0.62 Å) with $Al^{3+}$ ($R$=0.535 Å) and/or $Sc^{3+}$ ($R$=0.745 Å) ions[20]. The substantial differences in ionic radii enable a considerable shift in the phase transition temperature, even at low dopant concentrations, thereby extending the usable thermal range of the material. Moreover, our results demonstrate that the synthesis route employed for preparing $LaGaO_3$:$Eu^{3+}$ critically influences its thermometric performance. Specifically, controlling the morphology and increasing the particle size leads to a more sharply defined phase transition, which enhances relative sensitivity and significantly narrows the thermal hysteresis loop of the *LIR*. This, in turn, reduces the uncertainty in temperature determination. Altogether, the results presented here represent a significant advancement in the rational design of high-performance luminescent thermometers tailored for specific application requirements.

## 2. Experimental Section

*Materials and synthesis*

The materials were synthesized by using two different routes: conventional high-temperature solid-state reaction and the modified Pechini method[6,21].

*Solid-state synthesis of $LaGaO_3$:0.25%$Eu^{3+}$, $Al^{3+}$, $Sc^{3+}$*

The $LaGaO_3$:0.25% $Eu^{3+}$, $Al^{3+}$, $Sc^{3+}$, phosphors were prepared by using conventional high-temperature solid-state reaction method. $La_2O_3$ (99.999% purity, Stanford Materials Corporations), $Ga_2O_3$ (99.999% purity, Alfa Aesar), $Al_2O_3$ (99.995% purity, Alfa Aesar), $Sc_2O_3$ (99.99% purity, Alfa Aesar), $Eu_2O_3$ (99.99% purity, Stanford Materials Corporations) were used as starting materials. The raw materials were weighed in a stoichiometric ratio according to the required chemical formula, then mixed and ground in an agate mortar with a few drops of hexane for 30 minutes to ensure homogeneity. The resulting mixtures were annealed in corundum crucibles at 873 K for 3 hours in a muffle furnace, with a heating rate of 10 K min$^{-1}$. After cooling to room temperature, the powders were reground and sintered at 1673 K for 6 hours under the same heating rate. The final powders were then ground again in an agate mortar before further characterization.

*Modified Pechini synthesis of $LaGaO_3$:0.25%$Eu^{3+}$*

The $LaGaO_3$:0.25% $Eu^{3+}$ were synthesized via modified Pechini method. $Eu_2O_3$ (99.99% purity, Stanford Materials Corporations), $La_2O_3$ (99.999% purity, Stanford Materials Corporations), $Ga(NO_3)_3·9H_2O$ ( 99.999% purity, Alfa Aesar), $C_6H_8O_7$ (CA, >99.5% purity, Alfa Aesar) and $H(OCH_2CH_2)_nOH$, (PEG-200, n = 200, Alfa Aesar) were used as starting materials. A stoichiometric amount of La$_2$O$_3$ and Eu$_2$O$_3$ were dissolved in distilled water with the addition of 3 ml of nitric acid (65% solution, Avantor). The solution was then recrystallized three times to remove excess of nitric acid. The Ga(NO$_3$)$_3$·9H$_2$O was added to the water solution of lanthanide nitrates after recrystallization. After that, CA and PEG-200 were added to the mixture. The molar ratio of citric acid to all metal ions was set up as 6:1, meanwhile the PEG-200 and citric acid was used in a molar ratio of 1:1. The prepared solution was subsequently dried at 363 K for three days until a resin was formed. The produced resin was annealed in corundum crucibles under the same conditions as in the case of the solid-state reaction.

*Methods*

The purity of obtained materials was assessed through powder X-ray diffraction (XRD) technique. X-ray diffraction analysis of synthesized powders was conducted using a PANalytical X'Pert Pro diffractometer equipped with an Anton Paar HTK 1200N high-temperature stage employing Ni-filtered Cu Kα radiation (V=40 kV, I=30 mA). The measurements were carried out in 10 – 90º 2 θ range. ICSD database entries No.182539 (LT phase) and 182540 (HT Phase) were taken as initial models for the analysis of the obtained diffraction data.

The differential scanning calorimetry (DSC) measurements were carried out using a Perkin Elmer DSC 8000 calorimeter equipped with a Controlled Liquid Nitrogen Accessory LN2. The measurements were performed at a heating and cooling rate of 20 K min$^{-1}$, with a powder sample sealed in the aluminum pans.

The morphology and chemical composition of the obtained samples were assessed using a field emission scanning electron microscopy (FE-SEM, FEI Nova NanoSEM 230) integrated with an energy-dispersive X-ray spectrometer (EDX, EDAX Apollo X Silicon Drift Correction) compatible with genesis EDAX microanalysis Software. Before SEM imaging, the samples were dispersed in alcohol, and a drop of the resulting suspension was deposited onto a carbon stub. SEM images were subsequently acquired at an accelerating voltage of 5.0 kV using beam deceleration mode to enhance image quality. The estimation of the distribution of grain size was conducted through meticulous measurement of 100 grains utilizing the ImageJ software (version 1.53). For EDS analysis, measurements were acquired at an accelerating voltage of 30 kV.

The excitation and emission spectra were recorded using the FLS1000 fluorescence spectrometer (Edinburgh Instruments), equipped with a 450 W Xenon lamp as the excitation source and R928 photomultiplier tube (Hamamatsu) for signal detection. For temperature dependent measurements, the temperature of the sample was controlled by using a THMS600 heating-cooling stage from Linkam (0.1 K temperature stability and 0.1 K point resolution). The luminescence decay profiles were also acquired with the same spectrometer, utilizing a 150W µFlash lamp. The average lifetime of the excited state was determined by fitting decay curves using a double exponential function:

$$\qquad (1)$$

$$I(t) = I_0 + A_1 \cdot \exp\left(-\frac{t}{\tau_1}\right) + A_2 \cdot \exp\left(-\frac{t}{\tau_2}\right) \qquad (2)$$

where $\tau_1$ and $\tau_2$ are the decay components and $A_1$ and $A_2$ are the amplitudes of the double-exponential function.

For the proof-of-concept experiment, photographs were taken using a Canon EOS 400D camera equipped with an EFS 60 mm macro lens and a 450 nm long-pass optical filter (Thorlabs) to block the excitation beam. Luminescence images were acquired under UV excitation provided by a Vilber lamp ($\lambda_{exc}$ = 254 nm). Thermal treatment of the phosphors was performed on a laboratory hot plate, with the applied temperature additionally verified using a FLIR T540 thermographic camera, offering a measurement accuracy of ±0.5 K. The red and green channels (RGB) were extracted from the photographs using IrfanView 64 4.51 software. Subsequently, the R and G intensity maps were processed and divided by each other using ImageJ 1.8.0_172 software.

## 3. Results and discussion

The LaGaO$_3$ structure undergoes of a reversible first-order structural phase transition from the orthorhombic *Pbnm* (space group No. 62) phase to the rhombohedral *R3c* (No. 167) phase (Figure 1a), accompanied by a change in point symmetry from $C_S$ to $D_3$, respectively at around 430 K[15,16,18,22–25]. Such a transformation is particularly significant in systems doped with luminescent ions, which are highly sensitive to changes in the local symmetry and, consequently, exhibit modified luminescent properties due to occurring phase transitions. In this study, LaGaO$_3$ was doped with two types of ions. The first were luminescent ions - Eu$^{3+}$, which substitute La$^{3+}$ crystallographic positions and are coordinated by 8 to 12 oxygen atoms in distorted polyhedra. Beyond introducing Eu$^{3+}$ as the luminescent activator, the second coordination sphere was further engineered through the partial substitution of octahedrally coordinated Ga$^{3+}$ ions with Al$^{3+}$ and/or Sc$^{3+}$ ions. The most probable incorporation sites of these substituent ions were deduced based on the similarity of their ionic radii and their preferential coordination environments. The XRD analysis of the phosphors synthesized using solid state method confirmed the effective incorporation of both the dopant and substituent ions into the host lattice (Figure 1b, see also Figure S1, S2). The recorded diffractograms revealed no evidence of secondary phases or impurities across the full range of Al$^{3+}$ and Sc$^{3+}$ concentrations at a fixed Eu$^{3+}$ content of 0.25%, unequivocally confirming phase purity. However, the difference in the ionic radii between Ga$^{3+}$ and Al$^{3+}$ and Sc$^{3+}$ results in a shifting of the XRD reflections toward larger and smaller angles, respectively (Figure 1c). This effect results from the gradual shrinkage (for Al$^{3+}$) and expansion (Ga$^{3+}$) of the unit cell. Moreover, in the case of the phosphor doped with 10%$Al^{3+}$ and 15%$Al^{3+}$ instead of single reflection at around 32.6° two reflections around 32.5° and 32.7° occurs, that is an evidence of the formation of high temperature rhombohedral phase of the $LaGaO_3$ already at room temperature. The DSC curves analysis reveals that the first order phase transition occurs at around 430.1 K for $LaGaO_3$:0.25%$Eu^{3+}$ synthesized using solid state method (the influence of the synthesis condition on this aspect will be discussed in later part of this work) (see also Figure S3, S4).

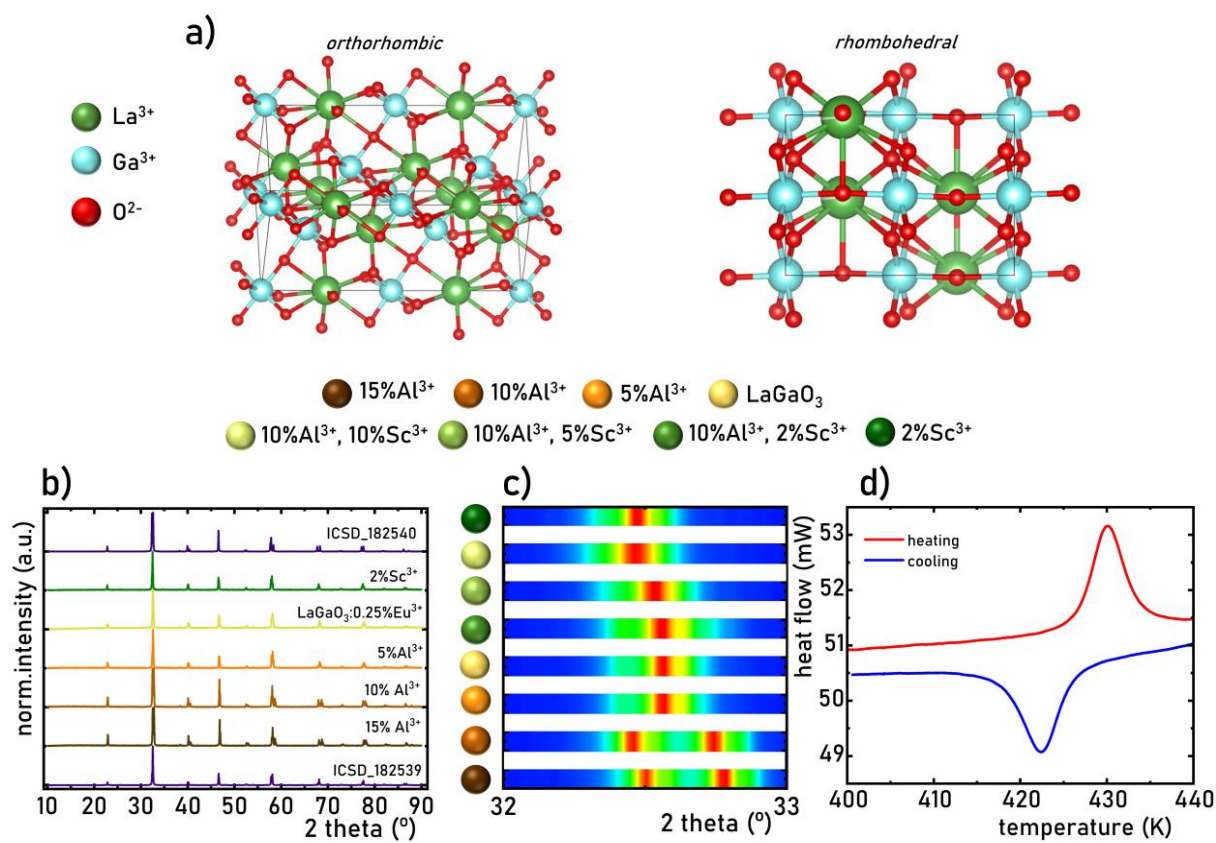

**Figure 1.** Visualization of the orthorhombic and rhombohedral $LaGaO_3$ structures -a) comparison of room temperature XRD pattern of representative $LaGaO_3$:0.25%$Eu^{3+}$, $Al^{3+}$,$Sc^{3+}$ phosphors synthesized using solid state method -b); comparison of the XRD patterns of all phosphors in the 32-33° 2theta range -c) DSC curves for $LaGaO_3$:0.25% $Eu^{3+}$ phosphor synthesized using solid state method -d)

The luminescence spectra of phosphors doped with $Eu^{3+}$ ions are primarily governed by radiative transitions from the excited $^5D_0$ level to the lower $^7F_J$ levels (Figure 2a)[26–29]. These electronic transitions give rise to characteristic emission bands typically observed at approximately 575 nm, 590 nm, 620 nm, 650 nm, and 700 nm, corresponding to the $^5D_0 \rightarrow {}^7F_0$, $^5D_0 \rightarrow {}^7F_1$, $^5D_0 \rightarrow {}^7F_2$, $^5D_0 \rightarrow {}^7F_3$, and $^5D_0 \rightarrow {}^7F_4$ transitions, respectively (Figure 2b). Due to the shielding of the *4f* orbitals by the outer *5s* and *5p* orbitals, the spectral positions of these emission bands remain relatively invariant across different host materials[30,31]. Among these transitions, the most intense are those corresponding to the $^5D_0 \rightarrow {}^7F_1$ and $^5D_0 \rightarrow {}^7F_2$, which are magnetic dipole and electric dipole transitions, respectively. The $^5D_0 \rightarrow {}^7F_2$ band, centered around 620 nm, is particularly sensitive to the local symmetry of the $Eu^{3+}$ site. In general, its intensity increases significantly with decreasing point symmetry, and it is typically weak in highly symmetric environments[26]. Each energy level of $Eu^{3+}$, except $^5D_0$ and $^7F_0$, is split into Stark components due to interactions with the crystal field of the host lattice. The number of Stark components increases both with decreasing site symmetry and with increasing *J* value[26]. Consequently, the spectroscopic properties of $Eu^{3+}$-doped phosphors are highly sensitive to the local crystallographic environment. This sensitivity makes $Eu^{3+}$ ions widely recognized in the literature as structural luminescent probes[27–29,32,33]. This strong dependence on the local structural environment renders $Eu^{3+}$ ions particularly well-suited for luminescence thermometry based on structural phase transitions. In the case of $LaGaO_3:Eu^{3+}$, the phase transition from orthorhombic to trigonal structure significantly alters the emission spectrum. A comparison of normalized emission spectra recorded at 83 K and 463 K (representative for low temperature (LT) and high temperature (HT), structural phases of $LaGaO_3$, respectively) for $LaGaO_3:0.25\%Eu^{3+}$ synthesized using solid state method reveals pronounced differences (Figure 2b). In the HT phase, the relative intensity of the $^5D_0 \rightarrow {}^7F_2$ transition increases with respect to the $^5D_0 \rightarrow {}^7F_1$ transition compared to the LT phase. Simultaneously, the relative intensity of the $^5D_0 \rightarrow {}^7F_4$ band decreases compared to $^5D_0 \rightarrow {}^7F_2$. Additionally, the shapes of these emission bands are noticeably altered after the phase transition. A detailed analysis of the three most prominent bands reveals that the increase in site symmetry accompanying the structural transition to the HT phase leads to a marked reduction in the number of Stark components (Figure 2 c-e). This is further accompanied by a spectral shift of the peak positions of Stark lines, attributed to changes in the local ions symmetry. These spectral modifications enable clear discrimination between the LT and HT phases, which is of critical importance for the thermometric applications discussed later in this work. Moreover, the structural phase transition in LaGaO$_3$:Eu$^{3+}$ also affects the radiative decay probability of the excited state, resulting in changes in luminescence decay kinetics. Specifically, the lifetime of the $^5D_0$ state varies from 1.06 ms to 0.81 ms for the LT to HT phases, respectively.

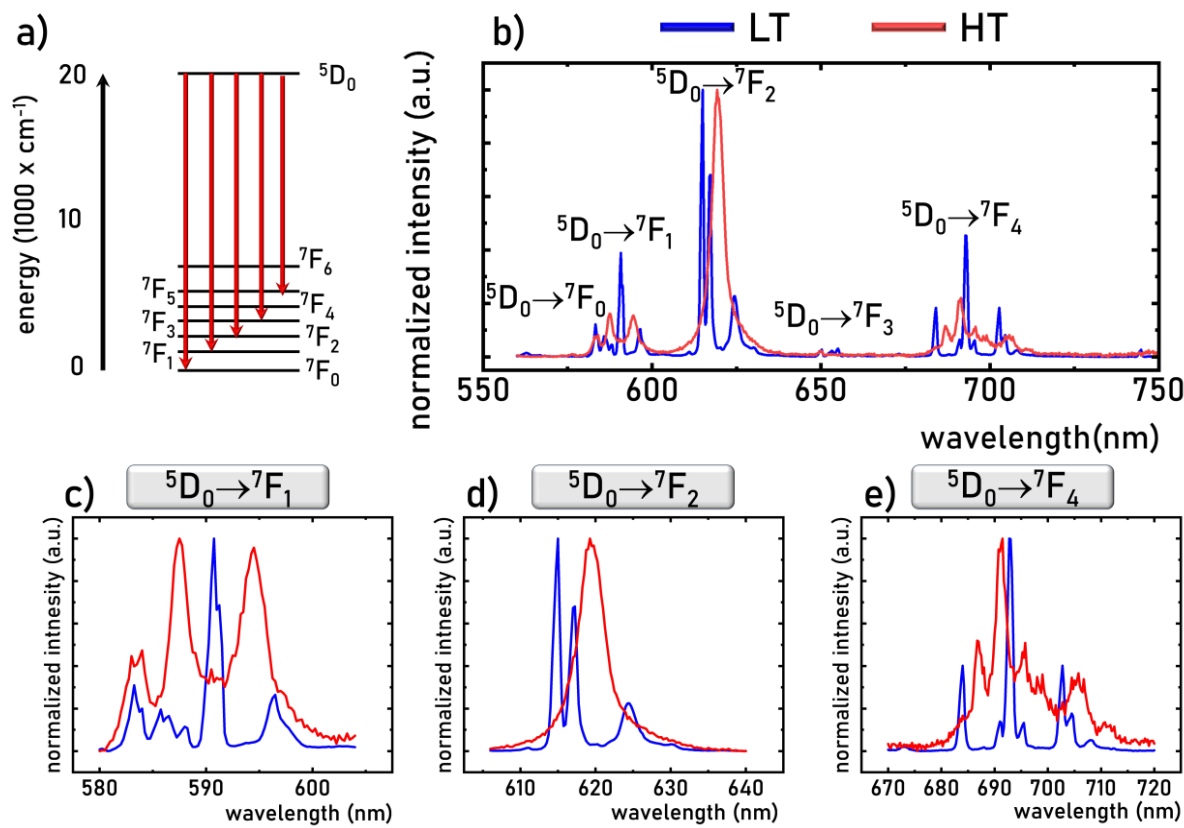

**Figure 2**. Simplified energy diagram of Eu$^{3+}$ ions -a) comparrison of normalized emission spectra of LaGaO$_3$:0.25%Eu$^{3+}$ measured and 83 K (blue curve) and 463 K (red curve) representative for LT and HT phases

of LaGaO$_3$:Eu$^{3+}$ respectively -b) and these spectra in the spectral range corresponding to the $^5D_0\rightarrow{}^7F_1$ -c),

$^5D_0\rightarrow{}^7F_2$ -d) and $^5D_0\rightarrow{}^7F_4$ -e) electronic transitions.

Although the thermometric performance of LaGaO$_3$:Eu$^{3+}$ has been previously reported in the literature[6], those studies have been limited to materials synthesized via the Pechini method. A characteristic feature of materials obtained through this method is a relatively broad distribution of crystallite sizes. Given that particle size can significantly influence the temperature at which a structural phase transition occurs, it may be expected that the choice of synthesis method may substantially affect the thermometric performance of luminescent thermometers based on this material. Therefore to verify this hypothesis, LaGaO$_3$ doped with 0.25% Eu$^{3+}$ was synthesized using both the Pechini method LaGaO$_3$:Eu$^{3+}$[P] and the conventional solid-state reaction method LaGaO$_3$:Eu$^{3+}$[SS]. Morphological analysis revealed clear differences between the two materials (Figure 3a-c, see also Figure S5-S8). The LaGaO$_3$:Eu$^{3+}$[P] exhibited small, strongly aggregated crystallites with an average grain size of 272 nm (Figure 3a, c), whereas the LaGaO$_3$:Eu$^{3+}$[SS] material consisted of significantly larger particles, averaging 634 nm (Figure 3b, c). The analysis of DSC curves of these two phosphors during the heating cycle confirmed the impact of morphology on the structural phase transition behavior (Figure 3d). The DSC curve of the LaGaO$_3$:Eu$^{3+}$[P] displayed a broader transition (FWHM = 18.18 K) compared to the sharper transition observed in the LaGaO$_3$:Eu$^{3+}$[SS] (FWHM = 4.68 K). Additionally, the peak transition temperature shifted from 432.28 K for the LaGaO$_3$:Eu$^{3+}$[P] to 430 K for the LaGaO$_3$:Eu$^{3+}$[SS] counterpart. The broader DSC curve in the former is attributed to overlapping of the input of phase transitions from particles of varying sizes. In contrast, the larger particle size achieved for LaGaO$_3$:Eu$^{3+}$[SS] leads to a more uniform temperature of phase transition across the sample. Emission spectra recorded as a function of temperature for both materials showed no major differences in spectral shape, aside from a slight narrowing of the Eu$^{3+}$ emission bands in the LaGaO$_3$:Eu$^{3+}$[SS] (Figure 3e, f). However, a marked difference was observed in thermal stability of their luminescence intensity. For LaGaO$_3$:Eu$^{3+}$[P], the luminescence intensity dropped to 25% of its initial value at around 400 K, while the LaGaO$_3$:Eu$^{3+}$[SS] retained this intensity level up to ~575 K (Figure 3g). This difference is attributed to the influence of particle size on thermal quenching behavior. Given the distinct spectral features associated with the LT and HT phases of LaGaO$_3$:Eu$^{3+}$, the material enables the development of a ratiometric luminescent thermometer. The thermometric parameter is defined as the luminescence intensity ratio (*LIR*) between selected emission bands as follows:

$$LIR_1 = \frac{Eu^{3+}(\text{rhombohedral})}{Eu^{3+}(\text{orthorombic})} = \frac{\int_{619nm}^{622nm}\left(^5D_0 \to {}^7F_2\right)d\lambda}{\int_{614nm}^{615nm}\left(^5D_0 \to {}^7F_2\right)d\lambda} \quad (3)$$

For both synthesis methods, the *LIR$_1$* increases gradually with temperature up to ~400 K, followed by a sharp rise and saturation of value near 550 K (Figure 3h). Upon further temperature increase, a decline in *LIR$_1$* is observed. Although the overall thermal *LIR$_1$* profile is qualitatively similar for both phosphors, notable differences exist. The threshold temperature at which the sharp *LIR$_1$* increase begins is lower for the LaGaO$_3$:Eu$^{3+}$[SS] (~410 K) in respect to the LaGaO$_3$:Eu$^{3+}$[P] (~450 K). Furthermore, the rate and magnitude of *LIR$_1$* increase above the threshold are higher in the LaGaO$_3$:Eu$^{3+}$[SS], which achieves the maximum *LIR$_1$* value over a narrower thermal interval. This more dynamic thermal response observed for LaGaO$_3$:Eu$^{3+}$[SS] is directly related to the narrower DSC curve, which in turn results from a larger crystallite size. To quantify the observed thermal dependences of *LIR$_1$* and thus to evaluate the thermometric performance of these thermometers, the relative sensitivity (*S$_R$*) was calculated as follows:

$$S_R = \frac{1}{LIR}\frac{\Delta LIR}{\Delta T} \cdot 100\% \quad (4)$$

The LaGaO$_3$:Eu$^{3+}$[P] exhibited a maximum $S_R$ ($S_{Rmax}$) of 3.0% K$^{-1}$ at ~460 K, while for the LaGaO$_3$:Eu$^{3+}$[SS] $S_{Rmax}$ = 18.2% K$^{-1}$ at ~420 K was reached (Figure 3i). This substantial difference clearly indicates the superior thermometric performance of LaGaO$_3$:Eu$^{3+}$ phosphors synthesized via the solid-state method, primarily due to enhanced thermal stability and sharper spectral transitions. As a result, the subsequent investigations in this study focus exclusively on materials synthesized using the solid-state method. Importantly, the observed effect is likely not unique to LaGaO$_3$:Eu$^{3+}$. It may also explain the discrepancies in $S_{Rmax}$ values previously reported in the literature for LiYO$_2$:Ln$^{3+}$ phosphors synthesized via the Pechini and solid-state methods[9,11,13].

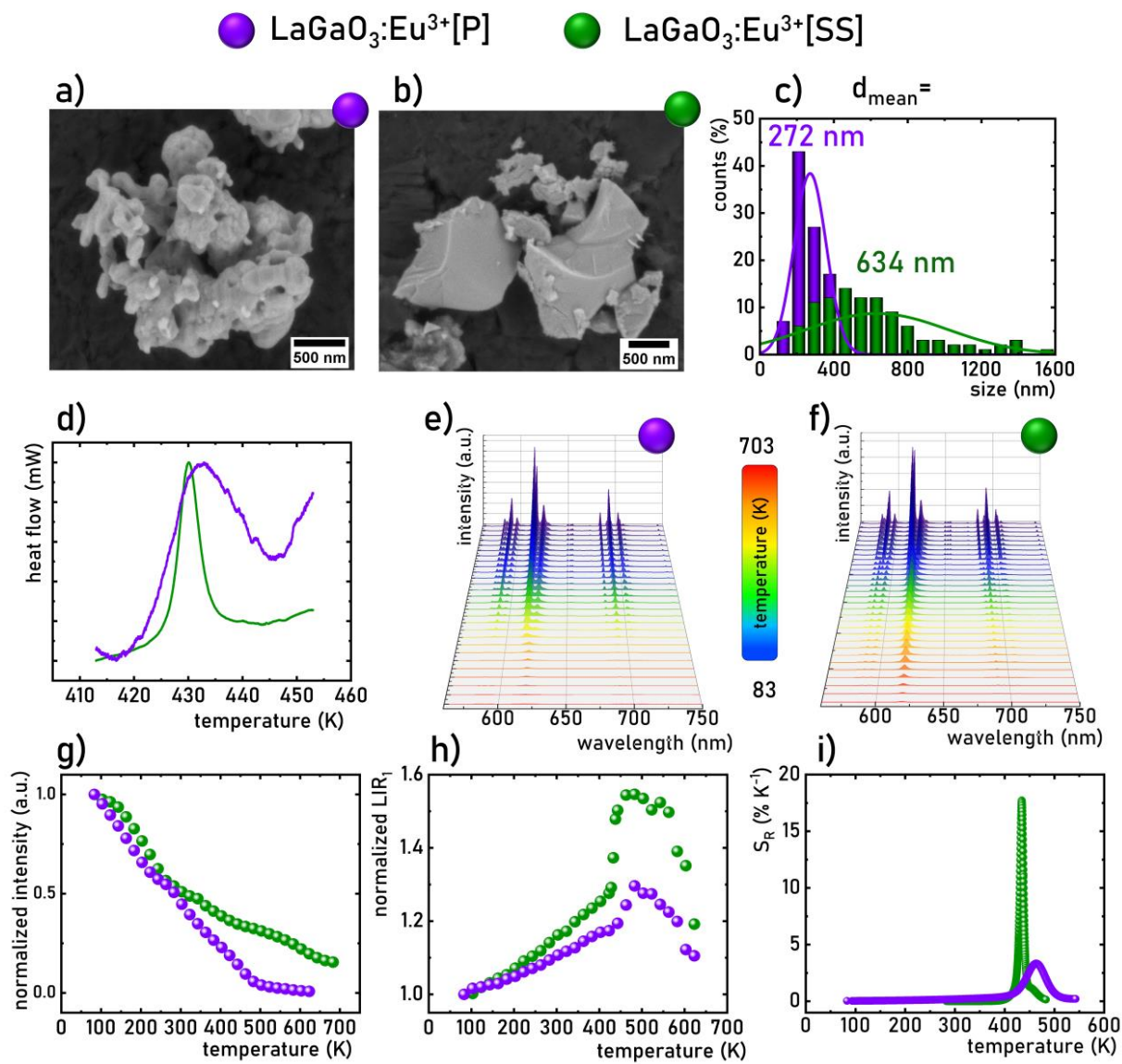

**Figure 3**. Representative SEM images of LaGaO$_3$:0.25%Eu$^{3+}$ synthesized using Pechini's-a) and solid state method -b) and corresponding particle size distribution -c); comparison of the DSC curves during heating cycle measured for this phosphor obtained using both methods-d); emission spectra of LaGaO$_3$:0.25%Eu$^{3+}$ synthesized using Pechini's method -e) and solid-state method -f) measured as a function of temperature; comparisson of thermal dependence of integral emission intensities of these phosphors -g); thermal dependence of normalized $LIR_1$ – h) and corresponding relative sensitivities -i).

One of the key limitations of luminescence thermometry based on thermally induced structural phase transitions is the relatively narrow thermal operating range, which is typically confined to the thermal vicinity of the phase transition temperature[12,13,34,35]. Hence, to tune a thermometric performance of such thermometers to the requirements of specific applications, it is essential to develop strategies that allow precise tuning of the thermal operating range. Recent studies have demonstrated that the introduction of co-dopants with ionic radii differing from those of the host cations can induce internal lattice strain, thereby shifting the phase transition temperature[8,13]. However, in systems where yttrium or other rare-earth (RE) cations are substituted with similarly sized RE ions, large dopant concentrations are often required to achieve only modest shifts in the transition temperature, due to the small differences in ionic radii.

In this context, the present study explores an alternative approach for LaGaO$_3$:Eu$^{3+}$, involving the substitution of Ga$^{3+}$ with Al$^{3+}$ and Sc$^{3+}$ ions while maintaining the phase purity of the material. Due to the more pronounced ionic radius differences between Ga$^{3+}$ and Al$^{3+}$/Sc$^{3+}$, a more significant modulation of the phase transition temperature is expected. In order to verify this hypothesis, the LaGaO$_3$:0.25%Eu$^{3+}$ co-doped with various concentrations of Al$^{3+}$, Sc$^{3+}$, and their combinations were synthesized. A comparison of their normalized emission spectra measured at 83 K revealed no significant changes in spectral shape, indicating minimal alteration in the local environment of Eu$^{3+}$ ions (Figure 4a, Figure S9-S15). This is consistent with the expectation that Al$^{3+}$ and Sc$^{3+}$, substituting Ga$^{3+}$, occupy positions in the second coordination sphere of Eu$^{3+}$ and thus limiting the influence on its immediate surroundings.

To quantitatively evaluate the structural mismatch introduced by dopant substitution, the $\Omega$ parameter, representing the relative mismatch in ionic radii, was calculated as follows:

$$\Omega = \frac{R_{Ga^{3+}} - R_{eff}}{R_{Ga^{3+}}} \qquad (5)$$

where $R_{Ga^{3+}}$ represents the ionic radii of Ga$^{3+}$ ions and effective ionic radii ($R_{eff}$) is defined as follows:

$$R_{eff} = \left(1 - x_{Sc^{3+}} - x_{Al^{3+}}\right) \cdot R_{Ga^{3+}} + x_{Al^{3+}} \cdot R_{Al^{3+}} + x_{Sc^{3+}} \cdot R_{Sc^{3+}} \tag{6}$$

As shown in Figure 4b, $\Omega$ varies from 0.02 for 15% $Al^{3+}$ ions to $\Omega=0.05$ for 2% $Sc^{3+}$. Intermediate values are observed for samples co-doped with both $Al^{3+}$ and $Sc^{3+}$, indicating that the $\Omega$ parameter can be finely tuned through precise control of dopant concentrations and host material composition. As it was already mentioned the intensity of the emission band attributed to the $^5D_0 \rightarrow {}^7F_2$ electronic transition is very sensitive to the change in the local $Eu^{3+}$ environment. Hence it is important to underline that the introduction of these co-dopants does not significantly alter the luminescence intensity ratio *LIR₂*, defined as the ratio of the $^5D_0 \rightarrow {}^7F_2$ to the $^5D_0 \rightarrow {}^7F_1$ transitions:

$$LIR_2 = \frac{\int_{610nm}^{630nm} \left({}^5D_0 \rightarrow {}^7F_2\right) d\lambda}{\int_{580nm}^{600nm} \left({}^5D_0 \rightarrow {}^7F_1\right) d\lambda} \tag{7}$$

Across all co-doped samples, *LIR₂* remains within the narrow range of 2.6–2.8, suggesting that the local symmetry around $Eu^{3+}$ ions remains largely unaffected (Figure 4c). This conclusion is further supported by time-resolved luminescence measurements. Decay profiles recorded at 83 K show nearly identical kinetics across all samples, regardless of the $Al^{3+}$ and $Sc^{3+}$ doping levels, confirming that the co-dopants do not significantly influence the radiative depopulation dynamics of $Eu^{3+}$ ions (Figure 4d). However, differences do emerge in the thermal dependence of the integrated luminescence intensity, reflecting variations in the phase transition temperature induced by the different co-dopant compositions (Figure 4e). These effects are analyzed and discussed in detail in the subsequent sections of this work.

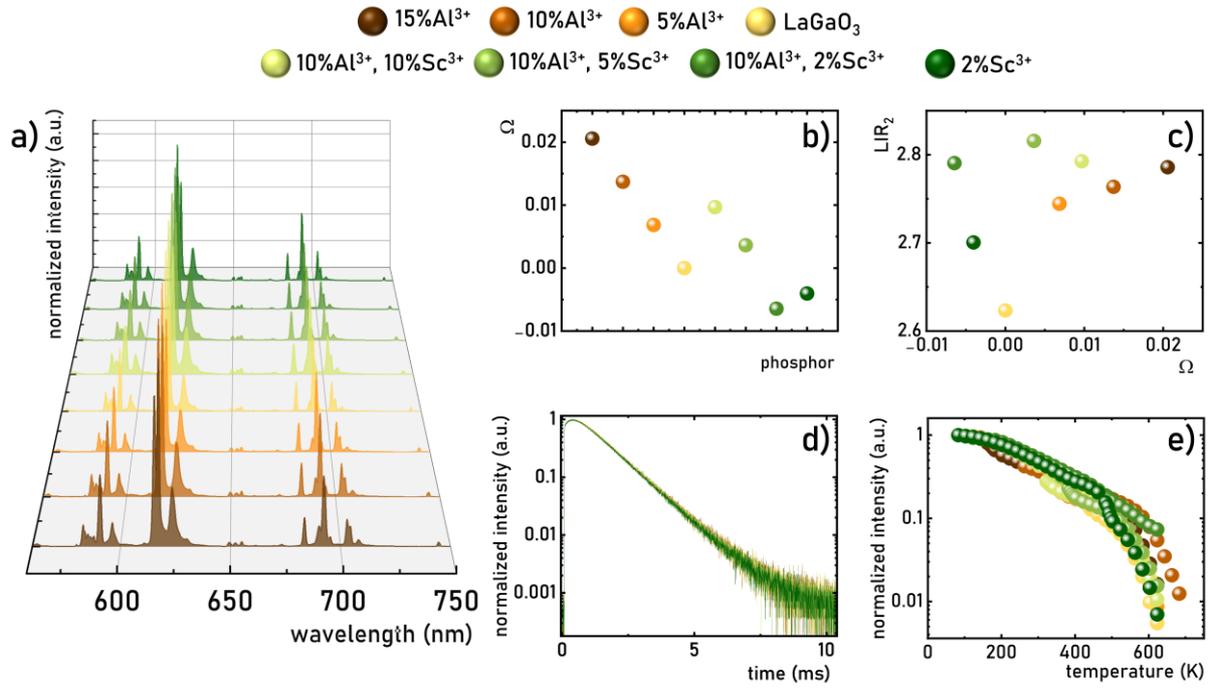

**Figure 4**. The comparison of emission spectra of LaGaO$_3$:0.25%Eu$^{3+}$,(Al$^{3+}$,Sc$^{3+}$) measured at 83 K -a), the influence of host material composition on $\Omega$ parameter -b); the influence of $\Omega$ on the *LIR$_1$* -c) comparisson of normalized luminescence decay curves ($\lambda_{exc}$=285, $\lambda_{em}$=620 nm) measured at 83 K for different host material composition -d); thermal dependence of integrated luminescence intensity -e) of LaGaO$_3$:0.25%Eu$^{3+}$,(Al$^{3+}$,Sc$^{3+}$).

As demonstrated by DSC measurements performed on LaGaO$_3$:Eu$^{3+}$ samples with varying concentrations of Al$^{3+}$ and Sc$^{3+}$, the structural phase transition temperature is highly dependent on the type and concentration of co-dopants. Specifically, for samples doped with 15% Al$^{3+}$, the transition temperature decreases to approximately 180 K, while a gradual reduction in Al$^{3+}$ content results in an increase in the phase transition temperature, reaching up to 420 K for LaGaO$_3$:0.25%Eu$^{3+}$. The introduction of 2% Sc$^{3+}$ further increases the transition temperature to approximately 500 K. Due to the fact that higher concentrations of Sc$^{3+}$ ions would result in even higher transition temperatures, potentially exceeding the measurement capabilities of our experimental setup, this study did not include samples with Sc$^{3+}$ concentrations beyond 2%. However, to verify the trend of increasing phase transition temperature with increasing Sc$^{3+}$ content, a series of samples co-doped with 10% Al$^{3+}$ and varying concentrations of $Sc^{3+}$ were synthesized. The presence of 10% $Al^{3+}$ ions effectively lowers the "baseline" transition temperature, thereby allowing observable changes induced by $Sc^{3+}$ incorporation to be captured within the accessible temperature range. These compositional effects are clearly reflected in the spectroscopic properties of the materials, particularly in the $^5D_0 \rightarrow ^7F_2$ emission band. Therefore, Figure 5a–h presents thermal luminescence maps of the normalized emission spectra focused on the spectral region corresponding to this electronic transition. The data reveal a distinct shift in the threshold temperature above which the HT phase of $LaGaO_3$:$Eu^{3+}$ is manifested. These spectral changes are directly mirrored in the thermal behavior of the luminescence intensity ratio $LIR_1$, as illustrated in Figure 5i. As the $Al^{3+}$ concentration decreases, the temperature range in which a sharp increase in $LIR_1$ occurs shifts to higher values. Similarly, the addition of 2% $Sc^{3+}$ increases the threshold temperature. The thermal evolution of $LIR_1$ is also reflected in the $S_R$ curves (Figure 5j). For the reference $LaGaO_3$:$Eu^{3+}$ sample, the maximum relative sensitivity $S_{Rmax}$ is observed at 440 K, while the introduction of $Al^{3+}$ and $Sc^{3+}$ co-dopants shifts the temperature range in which high $S_R$ values occur, albeit with a reduction in $S_{Rmax}$ values. To evaluate if this shift is associated with the ionic radii mismach, the temperature corresponding to $S_{Rmax}$ ($T@S_{Rmax}$) was plotted as a function of the $\Omega$ parameter, which quantifies the ionic radius mismatch between the host cations and the dopant ions (Figure 5k). The observed monotonic relationship between $T@S_{Rmax}$ and $\Omega$ confirms that the phase transition temperature, and consequently the temperature at which $S_{Rmax}$ is observed, is governed by ionic radius mismatch. Conversely, no clear correlation was observed between $S_{Rmax}$ and the $\Omega$ parameter (Figure 5l). As previously noted, all phosphor compositions exhibit a similar thermal profile for $LIR_1$, including a decline in $LIR_1$ beyond its saturation temperature. Since accurate temperature sensing requires a monotonic thermometric response over the entire operating range, the temperature at which $LIR_1$ begins to decrease effectively defines the upper limit of the thermal operating range ($T_{opr}$) for the luminescence thermometer. Importantly, this limiting temperature is tunable through appropriate control of co-dopant concentrations. For example, $T_{opr}$ increases from 83 - 223 K for 15% $Al^{3+}$ to 83 - 553 K for 2% $Sc^{3+}$, demonstrating the effectiveness of compositional tuning in optimizing the thermal operating window (Figure 5m).

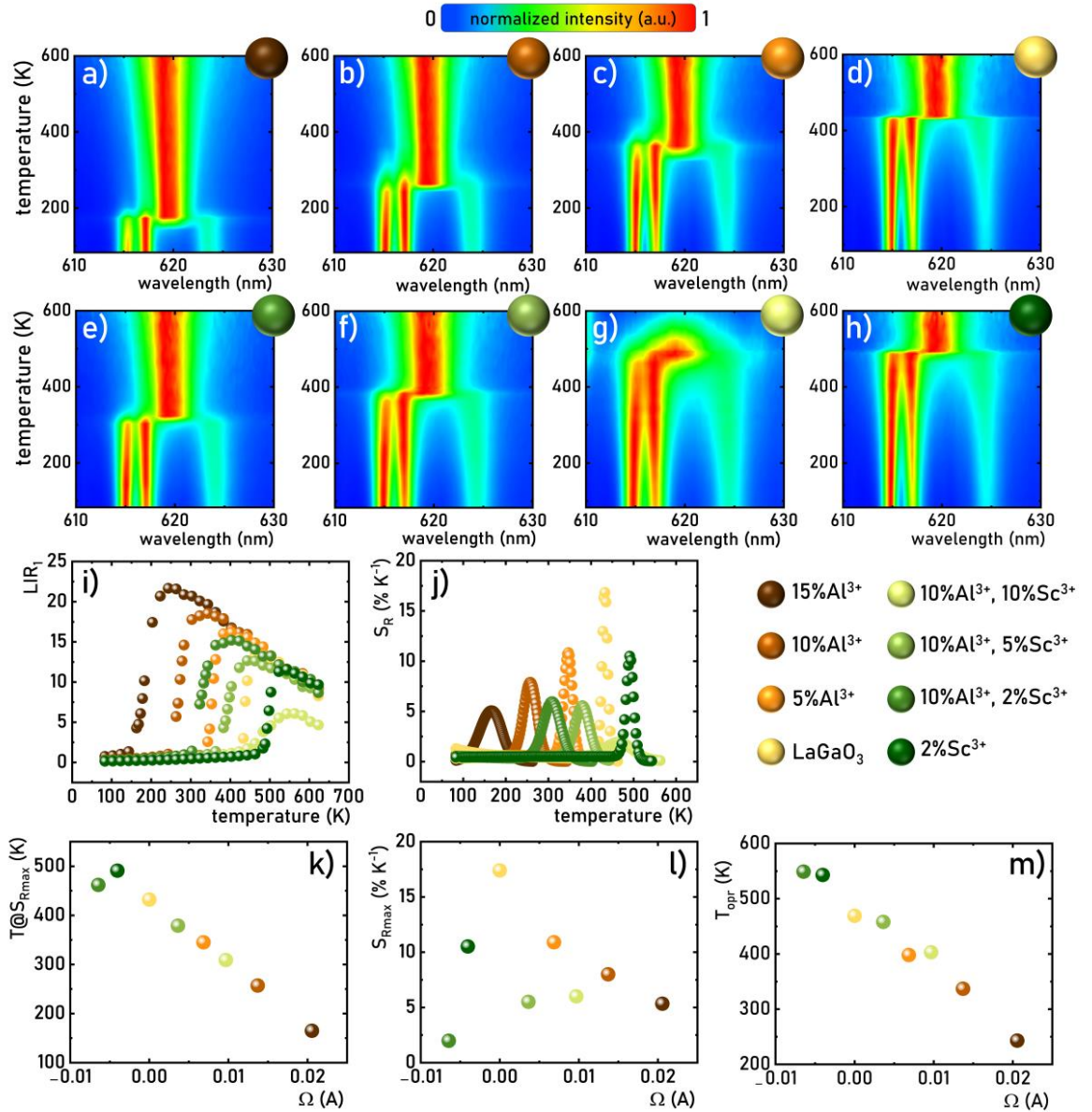

**Figure 5**. The normalized luminescence maps limited to the spectral range corresponding to the $^5D_0 \rightarrow ^7F_2$ emission band for LaGaO$_3$ co-doped with: 0.25%Eu$^{3+}$,15%Al$^{3+}$ - a), 0.25%Eu$^{3+}$,10%Al$^{3+}$ - b), 0.25%Eu$^{3+}$,5%Al$^{3+}$ - c), LaGaO$_3$:0.25%Eu$^{3+}$ - d), 0.25%Eu$^{3+}$, 10%Al$^{3+}$, 10%Sc$^{3+}$ - e); 0.25%Eu$^{3+}$,10%Al$^{3+}$, 5%Sc$^{3+}$ - f), 10%Al$^{3+}$,

2%Sc$^{3+}$ - g) and 0.25%Eu$^{3+}$, 2%Sc$^{3+}$ - h); thermal dependence of $LIR_2$ for different host material compositions – i); and corresponding $S_R$ – j); the influence of Ω on $T@S_{Rmax}$ -k) and $S_{Rmax}$ – l); the influence of Ω on $T_{opr}$ -m).

As noted in the Introduction, a key limitation of luminescent thermometers based on structural phase transitions is the thermal hysteresis of the thermometric parameter, which originates from the hysteresis behavior typically observed in DSC measurements. This hysteresis can introduce uncertainty in temperature determination (δT), particularly near the phase transition region. In the case of LaGaO$_3$:Eu$^{3+}$[P] and LaGaO$_3$:Eu$^{3+}$[SS] phosphors thermal hysteresis in the $LIR_1$ is clearly observed (Figure 6a, b, Figure S16-S22). However, the width of this hysteresis loop, defined as the difference in $LIR_1$ values between heating and cooling cycles, is significantly larger for LaGaO$_3$:Eu$^{3+}$[P] in respect to the one observed for LaGaO$_3$:Eu$^{3+}$[SS]. To quantify this effect, the differential luminescence intensity ratio (Δ$LIR_1$) was calculated as follows:

$$\Delta LIR_1(T) = LIR_1(T)_{cooling} - LIR_1(T)_{heating} \qquad (8)$$

$$\delta T = T\left[LIR_1(T)_{cooling} - LIR_1(T)_{heating}\right] \qquad (9)$$

The resulting data indicate that the largest discrepancies occur near the phase transition temperature, which is consistent with the expected behavior of first-order phase transitions (Figure 6c). For LaGaO$_3$:Eu$^{3+}$[P], the maximum Δ$LIR_1$ reaches 1.15, whereas for LaGaO$_3$:Eu$^{3+}$[SS], the maximum value in the same temperature range is only 0.5. These differences correspond to a temperature determination uncertainty (δT) of approximately 5.8 K for LaGaO$_3$:Eu$^{3+}$[P] and only 1.2 K for LaGaO$_3$:Eu$^{3+}$[SS] (Figure 6d). A similar analysis performed on other phosphors investigated in this study shows that Δ$LIR_1$ increases with higher concentrations of Al$^{3+}$ or Sc$^{3+}$, reaching a maximum of 0.8 for a 10% Al$^{3+}$ dopant level.

Importantly, this effect can be mitigated through co-doping with both $Al^{3+}$ and $Sc^{3+}$ (Figure 6e). The resulting temperature uncertainty is reduced to approximately 3 K for 10% $Al^{3+}$ and 2.2 K for 2% $Sc^{3+}$. Although a temperature uncertainty (δT) of 1.2 K may be too high for certain high-precision applications, it remains acceptable for others (Figure 6f). Furthermore, as demonstrated in previous studies, using the average *LIR* (i.e., the arithmetic mean of values obtained during heating and cooling) as the thermometric parameter can reduce the uncertainty by approximately 50%, offering an effective strategy to improve measurement reliability in systems exhibiting hysteresis[10].

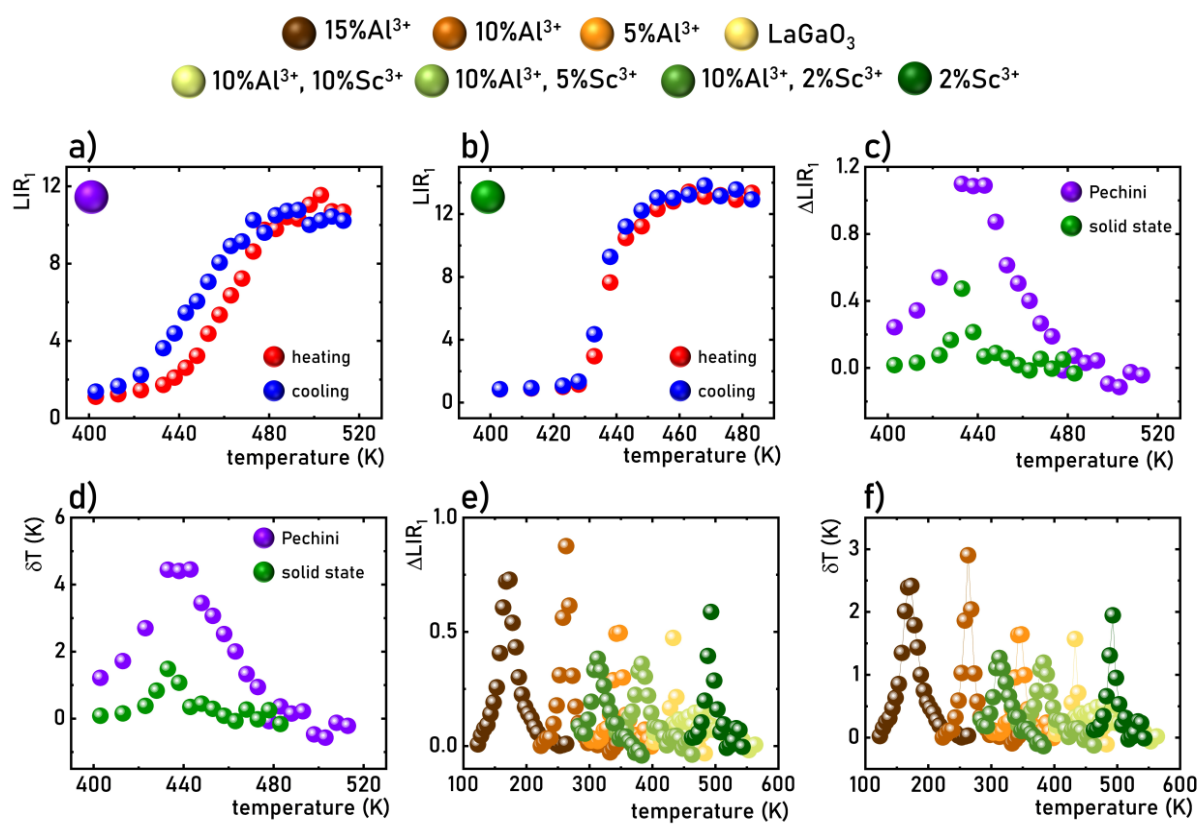

**Figure 6**. Thermal dependence of $LIR_1$ of $LaGaO_3$:0.25%$Eu^{3+}$ synthesized via Pechini method measured within heating-cooling cycle -a) and corresponding data for sample synthesized by solid-state method -b); thermal dependence of $\Delta LIR_1$ for these both phosphos -c); and corresponding *δT* -d); thermal dependence of $\Delta LIR_1$ for different host material composition -e) and corresponding *δT* – f).

As previously demonstrated, the phase transition temperature ($T_{PT}$) in LaGaO$_3$ exhibits a monotonic dependence on the $\Omega$ parameter, enabling the construction of an empirical function describing this relationship (Figure 7a). Although the correlation initially appears linear, a more accurate fit was obtained using a second-degree polynomial, resulting in a quadratic dependence described by the following form:

$$T_{PT}(\Omega) = 28165.03 \cdot \Omega^2 - 13501.21 \cdot \Omega^2 + 433.71 \qquad (10)$$

This relationship enables the extrapolation of the phase transition temperature based on the concentration of Al$^{3+}$ and/or Sc$^{3+}$ ions. Due to the smaller ionic radius of Al$^{3+}$ relative to Ga$^{3+}$, the effective ionic radius ($R_{eff}$) decreases with increasing Al$^{3+}$ content, reaching 0.535 Å for 100% Al$^{3+}$ substitution (LaAlO$_3$) (Figure 7b). In contrast, the considerably larger ionic radius of Sc$^{3+}$ results in an increase in $R_{eff}$ to approximately 0.745 Å upon full substitution (LaScO$_3$). Consequently, $\Omega$ increases up to 0.085 for Al$^{3+}$ doping, while for Sc$^{3+}$ it decreases to approximately -0.25 at 100% concentration (Figure 7c). Based on the derived quadratic relationship, it can be estimated that $T_{PT}$ is tunable from ~84 K (for 20% Al$^{3+}$) to ~1023 K (for 10% Sc$^{3+}$) (Figure 7d). While extrapolation to higher Sc$^{3+}$ contents must be approached cautiously without experimental validation, previous studies by Badie et al.[36] report the stability of the orthorhombic phase of LaScO$_3$ even at temperatures exceeding 1993 K. Although further investigation is needed to confirm the accuracy of the empirical model at higher dopant concentrations, the proposed methodology offers an effective strategy for controlling the phase transition temperature in LaGaO$_3$. This tunability enables the design and optimization of phase-transition-based luminescence thermometers across a broad temperature range of approximately 80-1023 K.

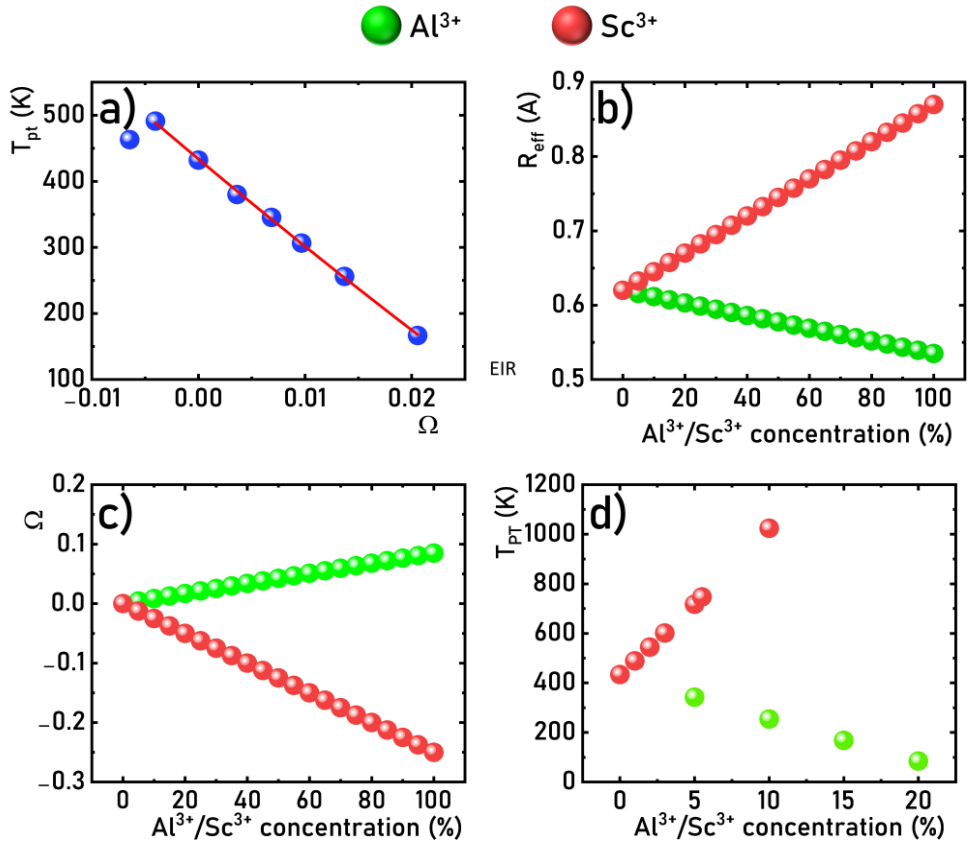

**Figure 7**. The influence of the $\Omega$ on the phase transition temperature ($T_{PT}$) – a); the influence of $Al^{3+}$ and $Sc^{3+}$ ions concentration of the effective ionic radii $R_{eff}$ – b), and on the $\Omega$ parameter – c); the calculated $T_{PT}$ as a function of $Al^{3+}$ and $Sc^{3+}$ ions concentration – d).

As demonstrated above, the LT to high-temperature HT structural phase transition in the analyzed materials results in a decrease in the intensity of the $^5D_0 \rightarrow {}^7F_1$ emission band relative to the $^5D_0 \rightarrow {}^7F_2$ transition, whereas upon cooling, the reverse behavior is observed. A comparison of normalized emission spectra of $LaGaO_3$:0.25%$Eu^{3+}$ in its LT and HT phases with the spectral sensitivity curves of the green (G) and red (R) channels of a conventional digital camera reveals that while the R channel encompasses all $Eu^{3+}$ emission bands, the G channel is predominantly sensitive to the $^5D_0 \rightarrow {}^7F_1$ transition (Figure 8a). Consequently, thermally induced spectral changes in these materials can be monitored by analyzing the ratio of images captured in the G and R channels. This method not only enables thermal imaging but significantly simplifies signal acquisition, as it does not require additional optical filters and utilizes only a standard digital camera. To validate the feasibility of this approach for thermal mapping using $LaGaO_3$:0.25%$Eu^{3+}$, $Sc^{3+}$/$Al^{3+}$ phosphors, a pattern composed of various phosphors was fabricated, as illustrated in Figure 8b (pattern diameter: 20 mm,). The phosphors were selected and arranged based on their differing phase transition temperatures, which correspond to different activation thresholds for green-channel luminescence enhancement. The patterned plate was placed on a hot stage at 453 K and excited using a UV lamp ($\lambda_{exc}$ = 254 nm) (Figure 8c, see also Figure S23). Luminescence images were recorded using a standard digital camera, and the G and R channel intensity maps were extracted. The ratio of these two channels (G/R) was calculated, producing a series of ratiometric images as the temperature was decreased to ambient conditions as shown in Figure 8d (Figure S24, S25). At 453 K, the pattern exhibited uniform red emission. Upon cooling, subtle color variations emerged, corresponding to the spatial distribution of phosphors with different transition temperatures. These variations became particularly pronounced in the G/R maps, where the progressive activation of green-channel emission in specific regions is clearly observable (Figure 8e, Figure S26). This experiment confirms the potential of phase-transition-based $Eu^{3+}$-doped phosphors not only for thermal imaging but also for applications in anti-counterfeiting technologies.

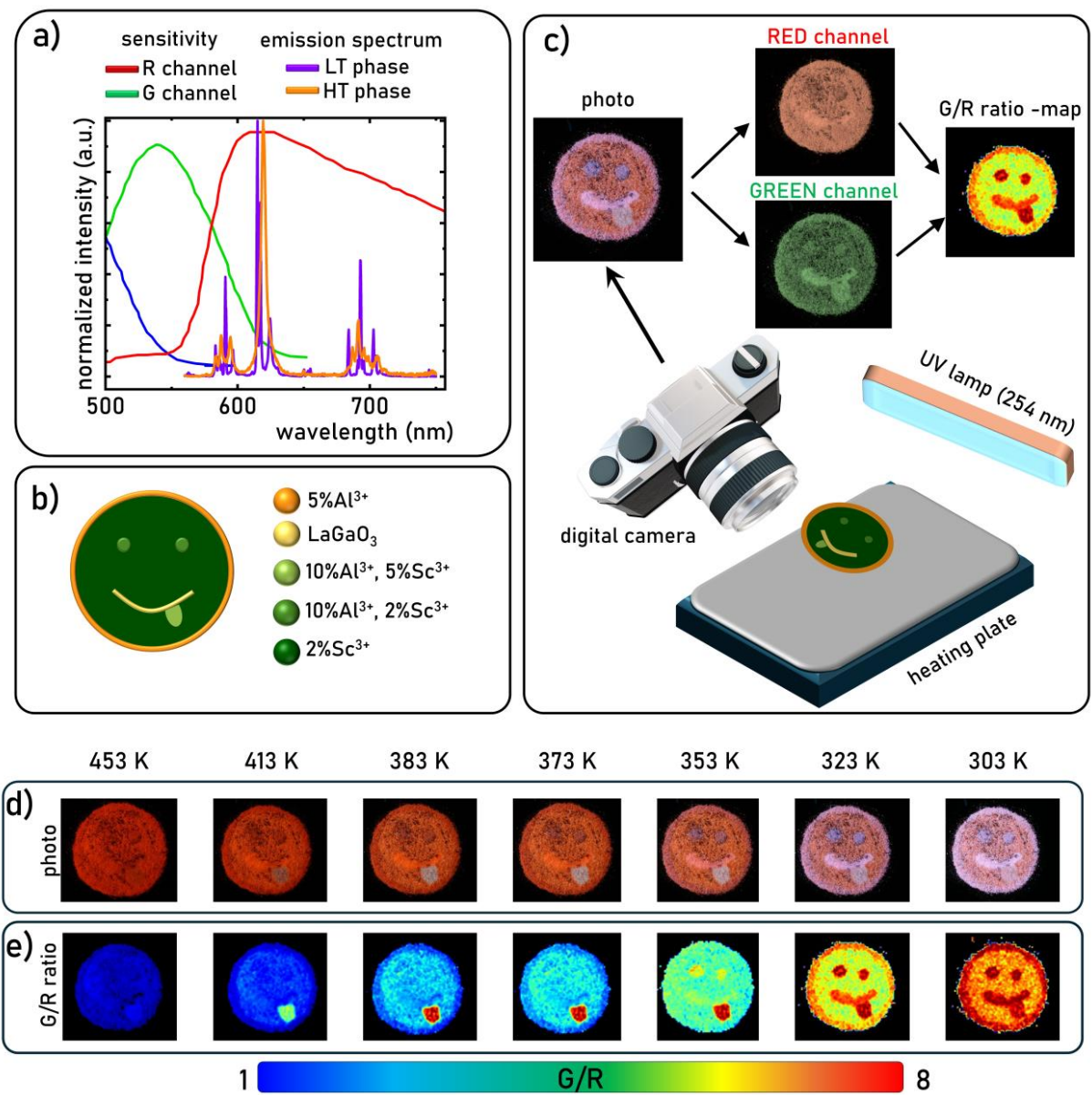

**Figure 8**. Normalized emission spectra of LT and HT phases of LaGaO$_3$:0.25%Eu$^{3+}$ comparred with the sensitivity curves of green and red chanels of digital camera -a); schematic representation of the phosphor aligment in the proof of concept experiment -b); experimental setup used in the experiment -c); representative photos of the luminescence of the prepared patterns obtained at different temperatures-d) and corrsponding maps of green to red ratio -e).

## Conclusions

In this study, the spectroscopic properties of LaGaO$_3$:Eu$^{3+}$ co-doped with Al$^{3+}$ and/or Sc$^{3+}$ ions were investigated as a function of temperature to evaluate their potential application in phase-transition-based ratiometric luminescent thermometry. As demonstrated, LaGaO$_3$ undergoes a structural phase transition above approximately 440 K, shifting from a low-temperature orthorhombic phase to a high-temperature trigonal phase. This structural transformation alters the local crystal field environment of the $Eu^{3+}$ ions, affecting both the relative intensities of emission bands and the number of Stark components into which the $Eu^{3+}$ energy levels are split. These temperature-dependent spectral changes enable the development of a ratiometric luminescent thermometer. It was also shown that the synthesis method significantly impacts the morphology of the resulting $LaGaO_3$:$Eu^{3+}$ phosphors. Specifically, materials synthesized via the Pechini method [P] exhibit smaller crystallites, while those obtained through the solid-state method [SS] show larger crystallites. This morphological difference is reflected in the thermal properties: broader peaks are observed in the DSC curves of $LaGaO_3$:$Eu^{3+}$[P] compared to the sharper, narrower peak for $LaGaO_3$:$Eu^{3+}$[SS]. As a result, the $LIR_1$ thermometric parameter exhibits a more abrupt thermal evolution in $LaGaO_3$:$Eu^{3+}$[SS], leading to substantially higher relative sensitivity values ($S_R$ = 18.2 % $K^{-1}$ for $LaGaO_3$:$Eu^{3+}$[SS] vs. $S_R$ = 3.0 % $K^{-1}$ for $LaGaO_3$:$Eu^{3+}$[P]). Moreover, the $LIR_1$ hysteresis loop is significantly narrower for $LaGaO_3$:$Eu^{3+}$[SS], resulting in a lower temperature determination uncertainty $\delta T$ of 1.2 K compared to $LaGaO_3$:$Eu^{3+}$[P], which exhibits $\delta T$ of 5.8 K. Spectroscopic analysis of emission spectra and luminescence decay profiles at 83 K indicated that substituting $Ga^{3+}$ with $Al^{3+}$ and/or $Sc^{3+}$ does not significantly alter the $Eu^{3+}$ emission characteristics, confirming that these dopants do not disrupt the immediate coordination environment of the $Eu^{3+}$ ions. However, their presence plays a critical role in tuning the structural phase transition temperature, which shifts from 180 K for 15% $Al^{3+}$ doping to 500 K for 2% $Sc^{3+}$ doping. This effect is attributed to the ionic radius mismatch between the dopant ions and the host cations, which introduces internal lattice stress, influencing the phase transition behavior. Due to the substantial differences in ionic radii, even low concentrations of $Al^{3+}$ and $Sc^{3+}$ are sufficient to induce significant changes in the transition temperature. Furthermore, the co-doping strategy allows for fine-tuning of the phase transition temperature and, consequently, the thermal operating range of the thermometer ranging from 83–180 K for 15% $Al^{3+}$ to 83–503 K for 2% $Sc^{3+}$. The observed monotonic relationship between the phase transition temperature and the relative ionic radius mismatch parameter ($\Omega$), along with the proposed empirical model describing this relationship, provides a robust foundation for the rational design of $LaGaO_3$:$Eu^{3+}$-based thermometers. This approach enables not only the controlled adjustment of the transition temperature but also the optimization of thermometric performance in luminescence thermometers exploiting structural phase transitions. The correlation between ionic radius mismatch and phase transition temperature enables precise tuning of the thermal response range, while the impact of synthesis-dependent grain morphology on thermal sensitivity and hysteresis underscores the importance of microstructural control. These findings provide a unified framework for designing luminescent thermometers with enhanced performance and structural responsiveness for advanced thermal sensing applications Moreover it was shown that phase transition-induced changes in the optical properties of $Eu^{3+}$ ions can be easily recorded by the digital camera by the analysis of G/R maps without the use of additional optical filters. The proper spatial arrangement of the these phosphors enables development of thermally activated patterns of high applicative potential in anticounterfeiting applications.

**Acknowledgements**

This work was supported by the National Science Center (NCN) Poland under project no. DEC-UMO- 2022/45/B/ST5/01629. M. Sz. gratefully acknowledges the support of the Foundation for Polish Science through the START program. L.M and M. Sz. would like to acknowledge support for Polish Academy of Science. L.T.K.G. would like to acknowledge support for Vietnam Academy of Science and Technology. under project No. QTPL01.02/24-25.